%% file: main.tex
\documentclass[titlepage, 10pt, a4paper, twocolumn]{article}
\usepackage{url}
\usepackage[utf8]{inputenc}
\usepackage[T1]{fontenc}
\usepackage{authblk}
\usepackage{cite}
\usepackage{graphicx}
\usepackage[none]{hyphenat}
\usepackage{tikz}
\usepackage{mathtools}
\usepackage{standalone}
\usepackage{etoolbox}

\makeatletter
\patchcmd{\thebibliography}{
  \chapter*{\bibname}\@mkboth{\MakeUppercase\bibname}{\MakeUppercase\bibname}}{
  \section{References}}{}{}
\makeatother

\twocolumn

\title{Predicting protein secondary structure with\\ Neural Machine Translation}
\author[1]{Evan Weissburg}
\author[2]{Ian Bulovic}
\affil[1,2]{Lexington High School}
\date{9/24/2018}

\begin{document}

\maketitle    

\section{Abstract \& keywords}{

We present analysis of a novel tool for protein secondary structure prediction using the recently-investigated Neural Machine Translation framework. The tool provides a fast and accurate folding prediction based on primary structure with subsecond prediction time even for batched inputs. We hypothesize that Neural Machine Translation can improve upon current predictive accuracy by better encoding complex relationships between nearby but non-adjacent amino acids. We overview our modifications to the framework in order to improve accuracy on protein sequences. We report 65.9\% Q3 accuracy and analyze the strengths and weaknesses of our predictive model.}
                        
\section{Introduction} \label{sec:Introduction}
Understanding protein secondary structure is critical for the study of protein folding and drug development. Secondary structure prediction has been studied since the early 1950s and is a bellweather of the growing bioinformatics field \cite{1950s}. Although current protein structure prediction techniques rely on fast sequence alignment, new approaches relying primarily on deep learning are proving to be more effective in producing high-fidelity predictions \cite{JPred}\cite{ConvNets}\cite{ComplexNN}.

We introduce a new model for secondary structure prediction based on the Neural Machine Translation (NMT) framework originally developed for natural language processing \cite{NMTOriginal}. Though NMT is best known for its use in Google Translate, we apply its unique encoding capabilities to the problem of protein structure prediction. We restate the problem of secondary structure prediction in terms of language, translating from the language of primary sequence to the language of secondary structure.

Both the primary and secondary structure of a given protein share the same underlying "meaning", characteristically similar to written language. This metaphor is well-formed since each "word", or amino acid, in the primary sequence corresponds directly to a "word", or state, in the secondary structure. We hypothesize that use of the NMT framework combined with these linguistic similarities will allow for increased predictive accuracy.

Most computational techniques predict secondary structure at the Q3 level, selecting either alpha-helix, beta-strand, or coil, while relatively few models predict at the 8-state level \cite{ConvNets}. Q8 prediction provides a greater amount of information about the overall structure but is equivalently a more difficult task. Additionally, there exist differing published mappings between 8-state and 3-state accuracy, complicating the conversion methodology \cite{Q3Q8Methods}\cite{Q3Q8Original}\cite{Q3Q8Method2}\cite{Q3Q8Method3}. Our model trains and reports accuracy at the Q8 level, but we report both 3-state and 8-state accuracy for comparison with literature.

Our implementation of the NMT framework consists of two LSTM-type Recurrent Neural Network cells: an encoder and a decoder \cite{LSTM}. The encoder processes the source protein in order, amino acid by amino acid, developing a "meaning vector", that encodes the protein's representation such that similar  proteins have similar meaning vectors. The decoder LSTM cell is initialized with the meaning vector, and uses this information to predict the most likely secondary structure, state by state \cite{NMTOriginal}.

The meaning vector is a powerful method for representing protein structure because of how well it encodes the relevant structure of a protein mathematically; as the NMT model trains, it optimizes its own encoding strategy \cite{HiddenState}. We note that not all proteins with similar primary structure have a similar meaning vector; certain amino acid substitutions may drastically change the local or global folding configurations of a protein. Human language exhibits this characteristic (e.g. "I am a doctor" vs "I am a patient") where sentences with similar structure often carry drastically different meanings \cite{NMT}. NMT systems are more effective than statistical natural language processing systems at translation-based tasks, and we expect this superiority to enable more comprehensive protein structure predictions \cite{NMTvsSMT}.

This article provides insight into the preprocessing, architecture, and metrics of our NMT model for secondary structure prediction, including a Github link\footnote{https://github.com/evanweissburg/tf-nmt} to the project. The repository contains ready-to-run scripts to download, cull, split, and fragment the requisite dataset from online. We explain two unusual mechanisms of our model added to improve the integrity of the meaning vector for protein structure prediction and overview commonly-used yet critical components of NMT models.

\section{Results}

We report 65.9\% Q3 (54.2\% Q8) validation accuracy using the hyperparameter set in Table \ref{tab:StandardHParams}. We will refer to these hyperparameters as standard for our work; deviations from the standard set will be specifically noted.

\input{hparams.tex}

We use Q3 and Q8 accuracy interchangeably for the purposes of illustrating different phenomena in the model; the two values are correlated with an R\textsuperscript{2} value of 0.995. Additionally, accuracy on protein fragments and entire protein sequences are similarly correlated, with an R\textsuperscript{2} value of 0.996. 

\subsection{Training Progress}

We present sample results for training, evaluation, and inference using the standard hyperparameter set. In each graph of performance as a function of training step number, trend lines are reverse-averaged for 40 time-steps.

\begin{figure}
  \center
  \includegraphics[width=\columnwidth]{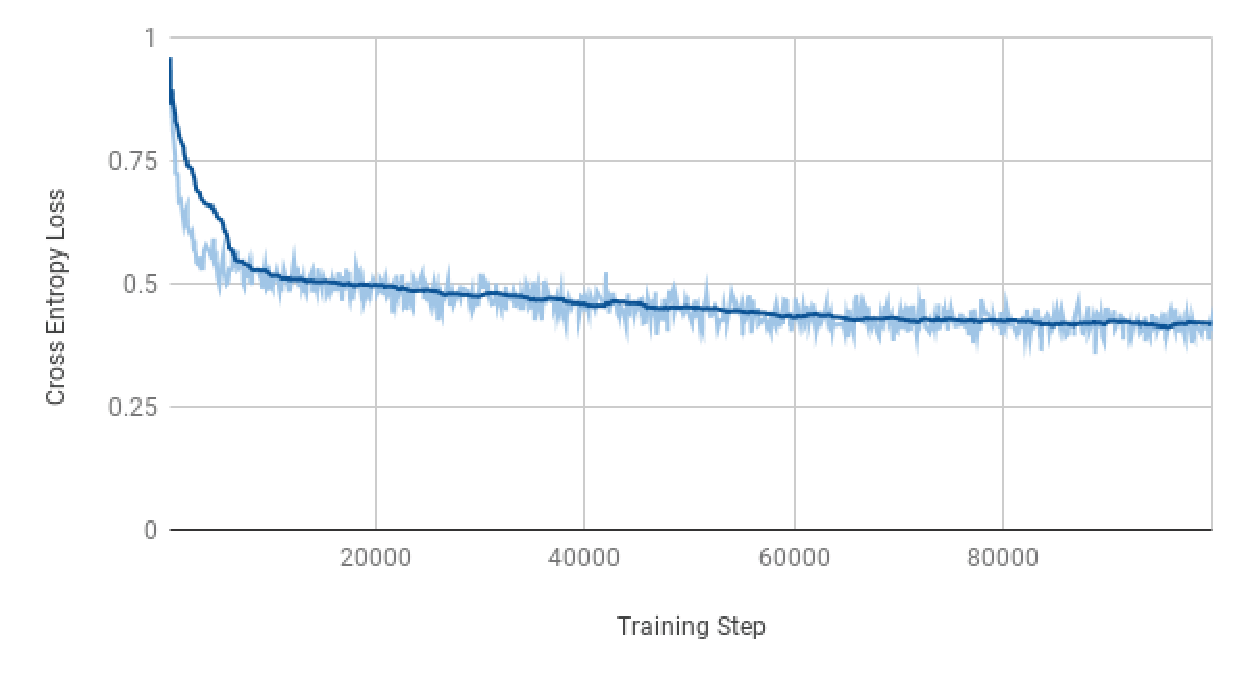}
  \caption{Training cross-entropy loss with the standard hyperparameter set using training dataset}
  \label{fig:ExampleTrain}
\end{figure}

Figure \ref{fig:ExampleTrain} illustrates training loss over training step, showing convergence to approximately $0.4$ cross-entropy loss.

\begin{figure}
  \center
  \includegraphics[width=\columnwidth]{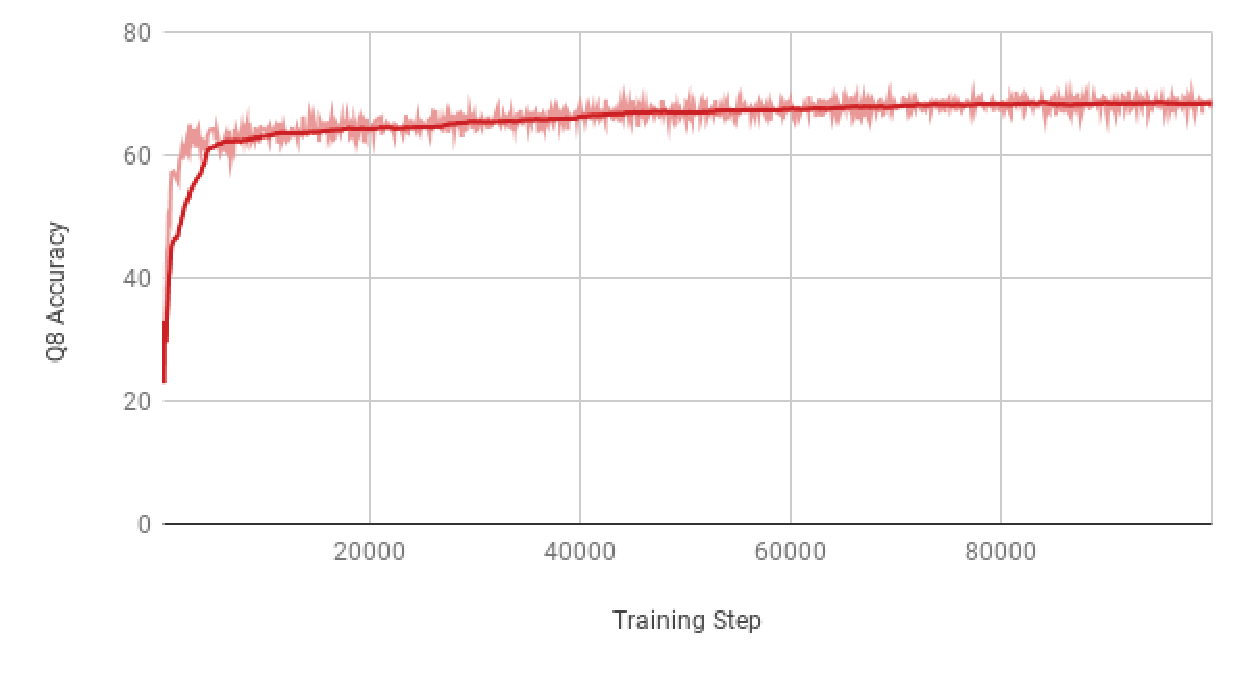}
  \caption{Evaluation 8-state accuracy with the standard hyperparameter set using testing dataset}
  \label{fig:ExampleEval}
\end{figure}

Figure \ref{fig:ExampleEval} illustrates evaluation 8-state accuracy over training step. NMT evaluation is a similar process to training, but it does not calculate gradient updates and uses a independent, non-overlapping dataset. Evaluation accuracy converges to approximately $67\%$, meaning that the model produces an accurate 8-state prediction 67\% of the time when provided the correct previous amino acid state.

\begin{figure}
  \center
  \includegraphics[width=\columnwidth]{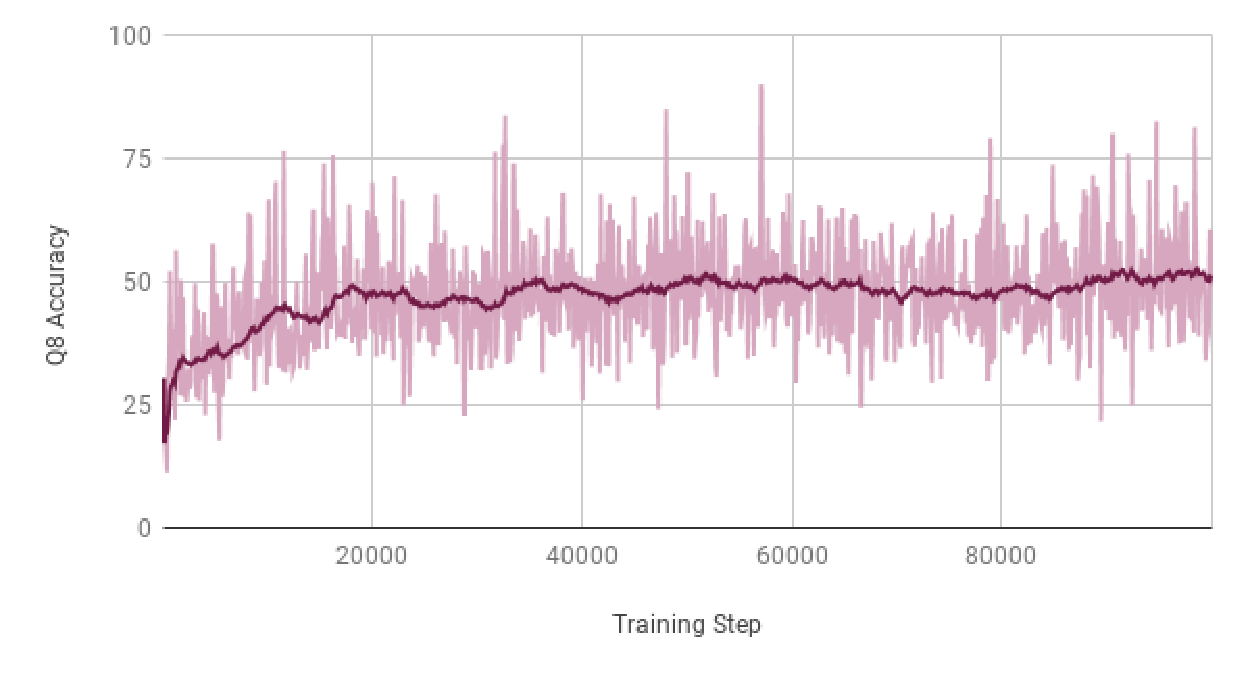}
  \caption{Inference 8-state accuracy with the standard hyperparameter set using testing dataset}
  \label{fig:ExampleInfer}
\end{figure}

Figure \ref{fig:ExampleInfer} illustrates inference 8-state accuracy over training step. Inference decoding differs from evaluation as the model is not provided the correct previous amino acid state, and must use its own previous predictions. Inference accuracy converges to approximately $50\%$.

\subsection{Confusion Matrix} \label{sec:confusion}

\begin{figure}
  \center
  \includegraphics[width=\columnwidth]{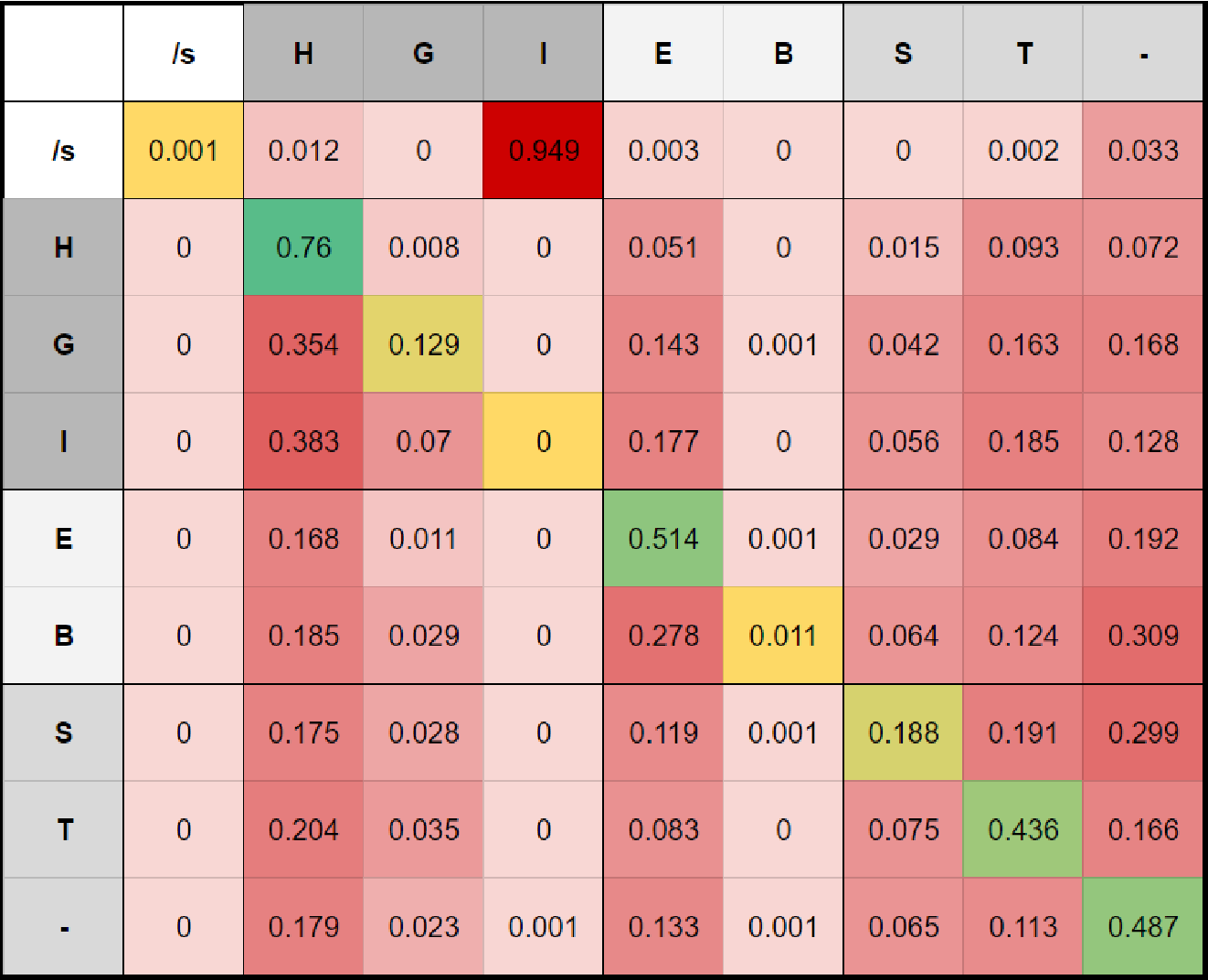}
  \caption{8-state confusion matrix displaying misprediction frequency per secondary state}
  \label{fig:ConfusionMatrix}
\end{figure}

The confusion matrix (Figure \ref{fig:ConfusionMatrix}) is based on validation accuracy using the standard hyperparameter set. Though the matrix is in terms of 8-state accuracy, we subdivide the matrix using black lines into 3-state classification for illustrative purposes.

The model is less effective at classifying lower frequency states, such as I (5-turn helix) or B (Isolated $\beta$-bridge). We also find a distinctly low false-negative rate for the most common state, H, showing evidence of class imbalance.

The end token (/s) is mispredicted as an I-type 94.9\% of the time, indicating that the model is generally unsure of when the protein sequence is completed. This behavior is expected since we only provide the model with fragments of a given protein.

H-type folds are the most common misprediction for nearly all fold types, due to mild class overrepresentation. Ignoring H-type mispredictions, we note that the most common misprediction is within the 3-state class two-thirds of the time. Coil (-) type folds are also a common misprediction for all states.

The matrix shows potential for improvement through increasing the number of LSTM layers, an approach that has found recent success in natural language processing but was not in the scope of this work \cite{NMT}. By using multiple stacked LSTM encoders and decoders, the model could maintain a more nuanced understanding of the underlying protein instead of simply storing more information in a single hidden state. This would hypothetically allow for higher accuracy, even on lower frequency states, such as G, I, or B.

\subsection {Hyperparameter Trials}

We explore the effects of variations to learning rate, network hidden size, and protein fragment radius on inference accuracy. We do not investigate batch size, since it has similar effects on accuracy as learning rate \cite{BatchSizeLR}. Source and target vocabulary sizes are not variable hyperparameters, since they correspond directly to the number of different amino acids and the number of Q8 states, respectively.

\subsubsection{Learning Rates}

\begin{figure}
    \center
    \includegraphics[width=\columnwidth]{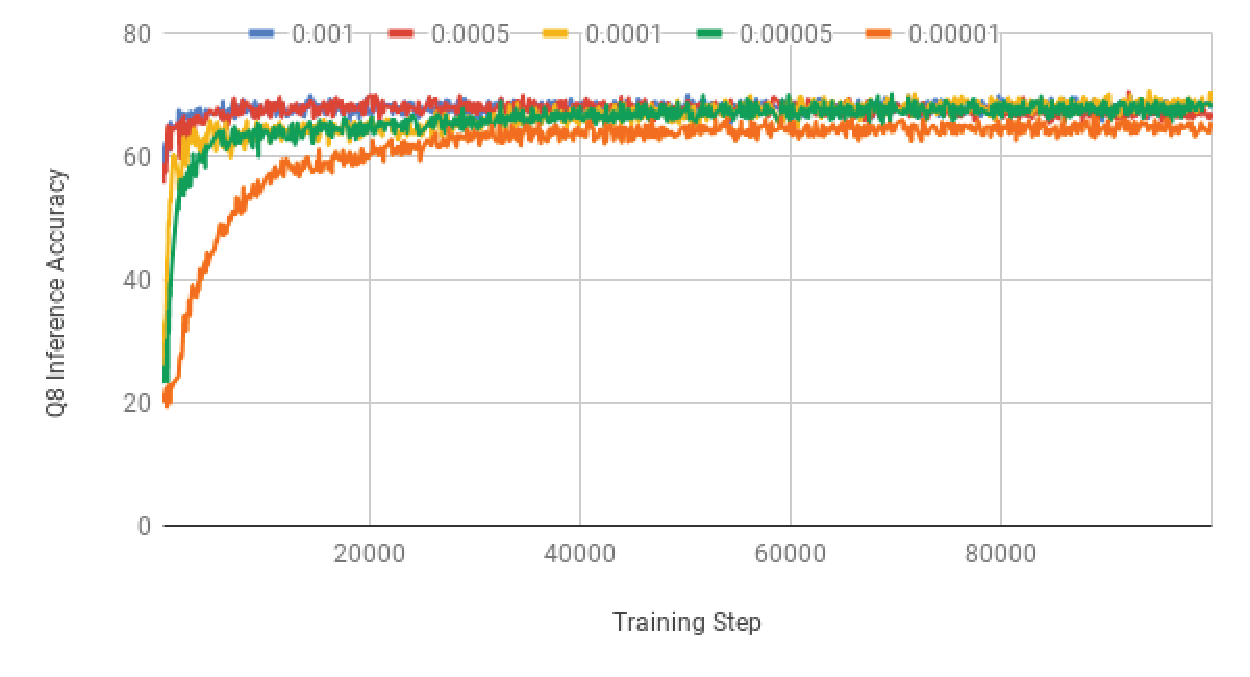}
    \caption{Effect of learning rate value on test-set evaluation Q8 accuracy}
    \label{fig:LearningRate}
\end{figure}

Figure \ref{fig:LearningRate} shows accuracy over time for different learning rate values. Although 0.0001 is generally accepted as a standard learning rate for neural translation models using Adam Optimizer, we explore the hyperparameter space more broadly due to our unique problem-space \cite{NMT}. We find that final inference accuracy is nearly independent from learning rate, which we attribute to the small size of our dataset; accuracy is resistant to changes in learning rate since the model does not get close enough to a local minimum of the error function regardless.

\subsubsection{Hidden Size and Fragment Radius}

\begin{figure}
    \center
    \includegraphics[width=\columnwidth]{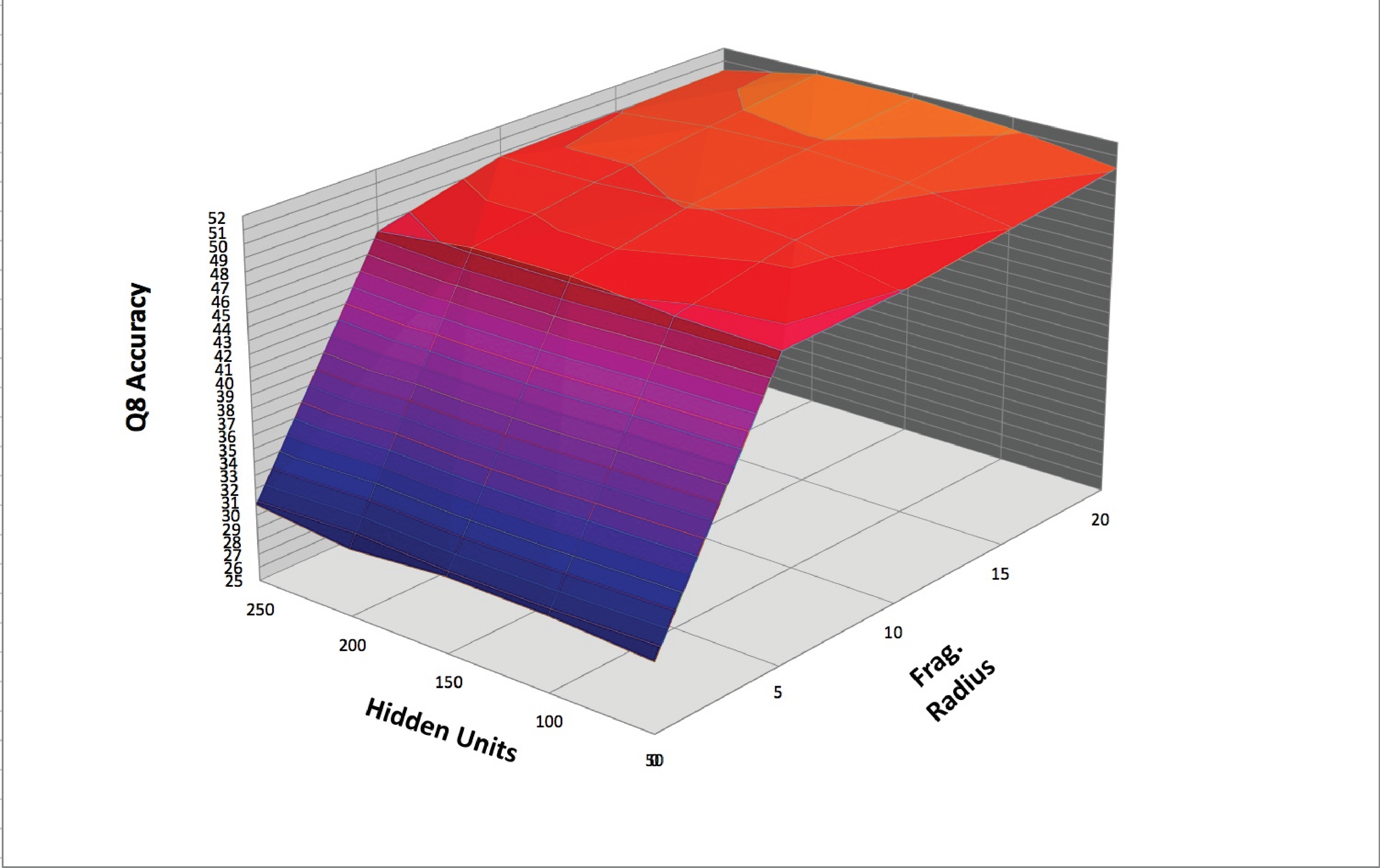}
    \caption{Effect of hidden size and fragment radius on Q8 inference accuracy}
    \label{fig:3DHParams}
\end{figure}

In Figure \ref{fig:3DHParams}, we show the joint effects of two hyperparameters on accuracy: network hidden size and protein fragment radius. Hidden size denotes the capacity of the meaning vector to store information, while fragment radius indicates the size of input protein fragment. The graph examines 20 total combinations; we test 5 hidden sizes (50, 100, 150, 200, 250) against 4 fragment radii (0, 5, 10, 20).

The graph illustrates an optimal network size of 150 hidden units. We observe overfitting with larger model sizes; evaluation accuracy diverges significantly earlier from training accuracy in models with 200 or more hidden units. 

The model attains ~30\% Q8 accuracy when trained with fragment radius 0 (single amino acids), consistent with the ~30\% frequency of the alpha-helix (H) state. Surprisingly then, the 0-radius model does not solely predict alpha-helix, tending towards the coil (-) prediction type when uncertain about an amino acid. A larger fragment radius correlates with higher accuracy, especially when the fragment radius is <10. Clearly, there exists an optimal fragment radius between 20 and 500 (the maximum length of a protein), where the trade-off between extra information and hidden state saturation is balanced; it appears to be effectively 20-30 based on the slope of the curve, which is consistent with the average number of words in a human sentence.

\section{Discussion}

We have presented a preliminary investigation into protein structure prediction using Neural Machine Translation. Although the model did not surpass modern architectures in either Q3 or Q8 accuracy, the application of a state-of-the-art framework to a novel use-case illuminates challenges of secondary structure prediction. First, relationships between amino acids differ in some ways from relationships between human words, since the same secondary state may appear in two completely different contexts. Although this phenomenon does occur in spoken language, words tend to have one or two meanings, while amino acid states frequently appear in millions of different, though related, contexts. We hypothesize that this problem is exacerbated in our model since, arguably, secondary structure confers less specific information about a protein than the original primary structure, unlike human language-to-language translation, where less information is lost.

Secondly, Neural Machine Translation is burdened by its requirement of a large dataset. Although there are over 300,000 proteins sequences in the Protein DataBank, only 17,000 sequenced protein have less than 20\% structural similarity \cite{CullPDB}. NMT systems are generally trained using millions of translation examples, so our model was hindered by the relatively small dataset currently usable for secondary structure prediction. Although small-scale NMT examples use 250-500 hidden units, we found that our dataset was not large enough to take best advantage of a complex meaning vector \cite{GNMT}.

The hidden units/fragment radius combined trials illustrate an exponential drop-off in accuracy past a fragment radius of 10. This curve is observed for all hidden units, though it is most pronounced near 150. This implies the meaning vector has become saturated; extra contextual information no longer aids the model's prediction. Normally, this problem is solved by increasing the size of the meaning vector, but any meaningful change in the number of hidden units tends to exacerbate overfitting in our implementation. 

It is difficult to compare the performance of our model fairly to existing architectures. State-of-the-art validation Q3 accuracy is about 84\%, while our initial implementation of NMT yields only 66\% \cite{ConvNets}. However, current approaches to secondary structure prediction are the result of years of iterative improvement, so it is perhaps more fair to compare our results to published explorations of novel architectures, where early implementations found validation accuracies between 50\% and 70\% \cite{Q3Q8Original} \cite{Early1} \cite{Early2}. Even this method of comparison is questionable; modern implementations of secondary structure prediction benefit from increased computational power and larger datasets.

Increasing the number of LSTM layers offers a potential next-step improvement to our results. As examined in section \ref{sec:confusion}, we expect to form a more nuanced hidden vector by using multiple connected LSTM cells. This approach would allow us to gain the benefits of increasing the hidden size while still minimizing overfitting.

\section{Methods}

We present prerequisite hardware and libraries to run and test our model. Additionally, we describe our dataset source, filtering, train/test split, and preprocessing algorithms. We present a mathematical overview of Neural Machine Translation and of the additional mechanisms added in our implementation. Finally, we overview the metrics and algorithms used to evaluate our results in context with literature.

\subsection{Hardware \& Prerequisites}

Training was performed on an Nvidia GTX 1070 graphics card, with 8 GB of VRAM memory.  Additionally, we note that checkpoint files can require up to 100 GB of disk storage.

Our NMT model was developed in Tensorflow\footnote{https://www.tensorflow.org/} with Python 3\footnote{https://www.python.org/download/releases/3.0/}, and has notable dependencies on the CUDA\footnote{https://developer.nvidia.com/cuda-downloads} and cuDNN\footnote{https://developer.nvidia.com/cudnn} libraries.

\subsection{Dataset \& Preprocessing}

\subsubsection{Source}

Data were obtained from the RCSB Protein Data Bank\footnote{https://www.rcsb.org/} and consists of more than 350,000 proteins, each with their primary and secondary structure annotated \cite{ProtDB}. The raw data describes primary structure in FASTA format, and secondary structure in DSSP 8-state format \cite{FASTA}\cite{DSSP}. Thus, the "vocabulary size" is 25 for the primary structure\footnote{24 FASTA characters + 1 <EOS> token} and 10 for the secondary structure\footnote{8 DSSP characters + 1 <SOS> token + 1 <EOS> token}. These small vocabulary sizes represent a major advantage for our implementation of NMT, since vocabulary size is generally the limiting factor in NMT training time and model size \cite{NMTVocabSize}.

\subsubsection{Filtering}

We limit structural similarity in our dataset by filtering by a list of proteins with no more than 30\% residual similarity provided by CullPDB \cite{CullPDB}. Additionally, we filter by length to use only proteins that have length <500 amino acids, eliminating a trivial portion of the dataset (<5\%). At this point, the dataset has been filtered to $17,000$ proteins.

\subsubsection{Train/Test/Validate Split}

We split the dataset into three independent sets, with 85\% ($\sim 14,000$ proteins) of the data being used for training, 10\% ($\sim 2,000$) for testing, and 5\% ($\sim 1,000$) for validation. Our model uses the train set to produce loss and gradients and the testing set to produce Q3/Q8 accuracy. The validation set is used only after model development is complete to compute reported Q3/Q8 accuracy.

\subsubsection{Fragmentation} \label{fragmentation}

We fragment each protein into numerous segments of shorter length. Since NMT models were originally developed for human sentences, we improve predictive accuracy by limiting the length of the protein fragments which the model receives.

The set of fragments $S$ is constituted by consecutive subsequences of the protein $p$, centered around amino acid $p_i$, each with "radius" $r$, such that

\begin{align*}
    & S_i = [p_{start}, ..., p_i, ..., p_{end}] \ \text{where}\\
    & \quad\quad start = max(i-r, 0)\\
    & \quad\quad end = min(i+r, length(p)-1)\\
    & \text{therefore} \quad length(S_i) <= 2r+1
\end{align*}

Once each protein fragment is processed, we combine the predictions to form a comprehensive secondary structure for the entire protein. We assume that each prediction fragment has highest accuracy in the center of the fragment (at position 11), since that state has the most context in terms of the number of amino acids surrounding it. Similarly, the first and last amino acid in the fragment will have the lowest degree of accuracy, since they lack the most positional context for their folding pattern. Therefore, when combining the predicted fragments to form a prediction for an entire protein, we weight each position in each fragment unequally, with amino acid states closer to the middle of the fragment receiving the most weight.

\begin{figure}
    \centering
    \includegraphics[width=\columnwidth]{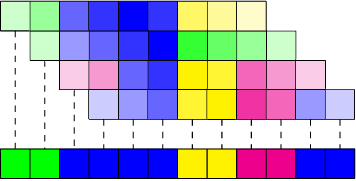}
    \caption{Example of prediction fragments being combined post-inference; color denotes 8-state secondary structure, opacity denotes weight of prediction}
    \label{fig:FragStitching}
\end{figure}

The final prediction is defined in terms of the fragment predictions and a vector of weights. The weights have the same length as each fragment and correspond to each position in each fragment, as discussed above. For each position in the final prediction, we calculate the weighted frequency of appearance of each state. The state with maximum frequency in the fragments at the current position is used in the final displayed prediction for the overall protein, as illustrated in Figure \ref{fig:FragStitching}.

\subsection{Model Overview}

\begin{figure}
  \center
  \includegraphics[width=\columnwidth]{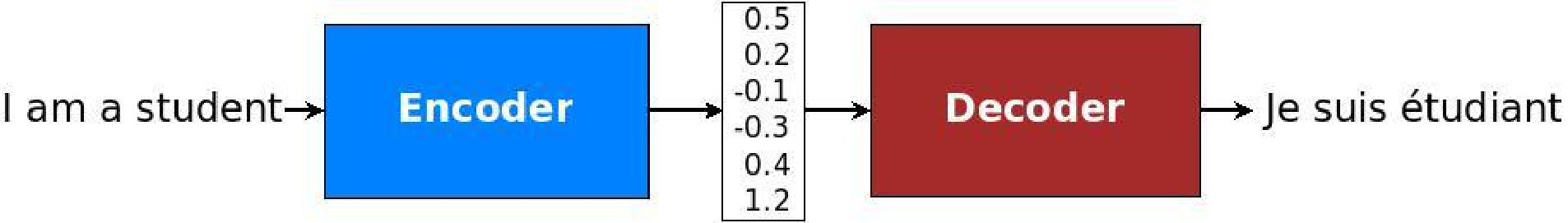}
  \caption{Encoder-decoder architecture \protect\cite{NMT}}
  \label{fig:EncDec}
\end{figure}

As overviewed in section \ref{sec:Introduction}, the NMT architecture consists of two LSTM-type recurrent neural network cells, an "encoder" and a "decoder", as Figure \ref{fig:EncDec} illustrates \cite{LSTM}. 

The encoder receives the input protein and produces a "meaning" vector that represents the protein mathematically. The decoder uses this universal vector representation to produce a high quality translation.

\subsubsection{Encoder}

The encoder cell receives the input protein $x$, amino acid by amino acid, recursively modifying the meaning vector $h$ such that \cite{GNMT}

\[h_t = f(x_t, h_{t-1})\]

\subsubsection{Decoder}

The decoder cell receives its previous prediction $y_{t-1}$, as well as the meaning vector $h$, which it continues to modify as it produces a prediction, such that

\[y_t, h_t = f(y_{t-1}, h_{t-1})\]

Thus, the probability of the next state is calculated based on the meaning vector $h$ and the previously predicted state $y$. The standard decoder is greedy, always choosing the state with the highest probability \cite{GNMT}.

We note that during training, the model receives the ground-truth correct previous prediction $y_{t-1}$ to improve training speed and accuracy \cite{NMT}. During inference, the model must use its previous prediction recursively, as illustrated above.

\subsubsection{Standard Mechanisms}

We implement the following mechanisms, standard in most modern NMT implementations \cite{NMT}.

\begin{enumerate}
  \item \textbf{Attention Mechanism}, allowing the decoder cell to access the stored meaning vector at different states as the encoder cell develops it \cite{NMT2014}.
  \item \textbf{Encoder Bi-directionality}, using two encoder cells instead of one. One encoder cell processes the primary structure normally, while the other processes it in reverse order. This allows for a better-formed meaning vector that remembers more of the source primary structure \cite{GNMT}.
  \item \textbf{Beam Search Decoding}, producing multiple candidate translations to compare at the end of decoding instead of greedily choosing the best state during decoding \cite{BeamSearch}.
\end{enumerate}

\subsubsection{Key Weights}

We introduce a modification to the standard NMT architecture to compensate for differences between human language and protein secondary structure. 8-state secondary folding is highly repetitive, with specific state classes occurring in segments on a protein. Therefore, the vanilla NMT decoder minimizes loss by repeating the previous prediction, without regard to the meaning vector, such that (wrongly)

\[y_t = y_{t-1}\]

We diminish this issue by re-weighting the loss function. Conventionally, each state is equally important to the overall prediction and the loss calculation. In our implementation, we weight positions in the structure prediction where the structure changes more highly than those where the state is the same as the previous state.

\begin{figure}
    \centering
    \includegraphics[width=\columnwidth]{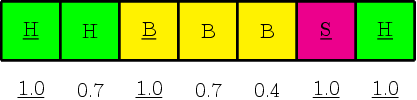}
    \caption{Example of the key weights system}
    \label{fig:KeyWeights}
\end{figure}

This modification allows the NMT system to focus on key points in the prediction, preventing it from falling into the local minimum explained above. The choice of weights adds additional hyper-parameters to the model; we chose a key point value of 1.0 and a weight decay value of 0.3, as in Figure \ref{fig:KeyWeights}.\\

\subsection{Metrics}

We assess training loss as well as testing/validation accuracy.

\subsubsection{Loss}

We compute cross-entropy loss $H$ for each predicted state to maximize log-likelihood of the training data, comparing the ground-truth probability distribution $P$ to the predicted probability distribution $Q$, such that \cite{NMT2014}

\[H(P,Q)=-\sum _{i}P[i]\log Q[i]\]

\subsubsection{Q3/Q8 Accuracy}

Q8 accuracy is computed by comparing the number of correct states to the total number of amino acids, yielding a raw percentage. This 8-state accuracy is converted to Q3 accuracy since each Q8 state maps to a distinct Q3 state by table \ref{tab:Q3Q8}, in accordance with the mapping originally proposed by Rost and Sander \cite{Q3Q8Original}\cite{Q3Q8Secondary}.

\input{3state8state.tex}

\section{Acknowledgements}

The authors thank Professor Bruce Tidor and Kevin Shi at MIT CSAIL for their guidance and insight during the research, drafting, and reviewing phases of the project. Our discussions with them were crucial for the development process. The authors declare no conflicts of interest.

\let\stdsection\section
\def\section*#1{\stdsection{#1}}

\bibliography{sample}{}
\bibliographystyle{ieeetr}

\let\section\stdsection

\end{document}

%% file: hparams.tex
\begin{table}
    \centering
    \begin{tabular}{c|c}
        Hyperparameter & Value \\
        \hline
        Learning Rate & 0.0001 \\
        Hidden Size & 150 \\
        Fragment Radius & 10 \\
        Batch Size & 500 \\
        Source Vocab. Size & 25 \\
        Target Vocab. Size & 10 \\
    \end{tabular}
    \caption{The standard hyperparameter set}
    \label{tab:StandardHParams}
\end{table}

%% file: 3state8state.tex
\begin{table}
    \centering
    \begin{tabular}{c|c}
        3-State & 8-State \\
        \hline
        Alpha Helix & H, G, I \\
        Beta Sheet & E, B \\
        Coil & S, T, -
    \end{tabular}
    \caption{Q3 to Q8 conversions}
    \label{tab:Q3Q8}
\end{table}